\documentstyle[12pt,epsf]{article}
\begin{document}

\font\fortssbx=cmssbx10 scaled \magstep2
\hbox to \hsize{
\hbox{\fortssbx University of Wisconsin - Madison}
\hfill$\vcenter{\hbox{\bf MADPH-98-1050}
                \hbox{April 1998}}$ }

\vskip.25in

\begin{center}
{\large\bf The AMANDA Neutrino Telescope and\\[1mm]
the Indirect Search for Dark Matter}\footnote{Talk presented at the {\it 3rd International Symposium on Sources and Detection of Dark Matter in the Universe (DM\,98)}, Santa Monica, California, Feb.~1998.}\\[2mm]
{\normalsize Francis Halzen, for the AMANDA Collaboration}\\[2mm]
\end{center}

\medskip

{\small
\begin{center}
R.C.~Bay,
Y.~He,
D.~Lowder,
P.~Miocinovic,
P.B.~Price,
M.~Solarz,
K.~Woschnagg\\
{\it University of California, Berkeley, USA}\\

\medskip

S.W.~Barwick,
Je.~Booth,
P.C.~Mock,
R.~Porrata,
E.~Schneider,
G.~Yodh\\
{\it University of California, Irvine, USA}\\

\medskip

D. Cowen\\
{\it University of Pennsylvania, USA}\\

\medskip

M.~Carlson,
C.G.S.~Costa,
T.~DeYoung
L.~Gray,
F.~Halzen,
R.~Hardtke,
J.~Jacobsen,
V.~Kankhadai,
A.~Karle,
I.~Liubarsky,
R.~Morse,
S.~Tilav\\
{\it University of Wisconsin, Madison, USA}\\

\medskip

T.C.~Miller\\
{\it Bartol Research Institute,USA}\\

\medskip

E.C.~Andr\'es,
P.~Askebjer,
L.~Bergstr\"om,
A.~Bouchta,
E.~Dalberg,
P.~Ekstr\"om,\\
A.~Goobar,
P.O.~Hulth,
C.~Walck\\
{\it Stockholm University, Sweden}\\

\medskip

A.~Hallgren,
C.~P.~de~los~Heros,
P.~Marciniewski,
H.~Rubinstein\\
{\it University of Uppsala, Sweden}\\

\medskip

S.~Carius,
P.~Lindahl\\
{\it Kalmar University, Sweden}\\

\medskip

A.~Biron,
S.~Hundertmark,
M.~Leuthold,
P.~Niessen,
C.~Spiering,
O.~Streicher,\\
T.~Thon,
C.H.~Wiebusch,
R.~Wischnewski\\
{\it DESY --- Institute for High Energy Physics, Germany}\\

\medskip

D.~Nygren\\
{\it Lawrence Berkeley National Laboratory, USA}\\

\medskip

A.~Jones,
S.~Hart,
D.~Potter,
G.~Hill,
R.~Schwarz\\
{\it South Pole Winter-Overs, Antarctica}\\

\end{center}}

\vskip.25in

\begin{abstract}
With an effective telescope area of order $10^4$~m$^2$, a threshold of $\sim$50~GeV and a pointing accuracy of 2.5~degrees, the AMANDA detector
represents the first of a new generation of high energy neutrino telescopes, reaching a scale envisaged over 25 years ago. We describe its performance, focussing on the
capability to detect halo dark matter particles via their annihilation into
neutrinos.
\end{abstract}

\thispagestyle{empty}
\newpage
\let\Large=\large
\section{The Indirect Detection of Halo Dark Matter}
\unskip

High energy neutrino telescopes are multi-purpose instruments; their science
mission covers particle physics, astronomy and astrophysics, cosmology and
cosmic ray physics. Their deployment creates new opportunities for glaciology
and oceanography, possibly geology of the earth's core\cite{pr}. Astronomy
with neutrinos does have definite advantages. They can reach us, essentially
without attenuation in flux, from the largest red-shifts. The sky is, in
contrast, partially opaque to high energy photons and protons because of
energy-loss suffered in interactions with infrared light, CMBR photons and
radio waves\cite{cronin}. They do not reach us from distances much larger than
tens of megaparsecs once their energy exceeds thresholds of 10~TeV for photons and $5\times 10^7$~TeV for protons. (Below this energy charged protons
do not point back to their sources.) The drawback is that neutrinos are difficult to detect: the small interaction cross sections that enable them to
travel without attenuation over a Hubble radius, are also the reason why
kilometer-scale detectors are required in order to capture them in sufficient
numbers to do astronomy\cite{halzenkm}. Some opportunities may, however, be
unique. If, for instance, the sources of the highest energy cosmic rays are
beyond $10^2$~Mpc, conventional astronomy is unlikely to discover them.

Some science missions do not require a detector of kilometer size. The best
opportunities to search for halo dark matter are, in fact, associated with the
present instrument which, while smaller in telescope area than the planned
extension of AMANDA to ICE3 (ICECUBE), has a lower threshold.

At this meeting, the capability of neutrino telescopes to discover the particles that constitute the dominant, cold component of the dark matter is
of special interest. The existence of the weakly interacting massive particles
(WIMPs) is inferred from observation of their annihilation products. Cold
dark matter particles annihilate into neutrinos; {\it massive} ones will
annihilate into {\it high-energy} neutrinos which can be detected in high-energy neutrino telescopes. This so-called indirect detection is greatly
facilitated by the fact that the earth and the sun represent dense, nearby
sources of accumulated cold dark matter particles. Galactic WIMPs, scattering
off nuclei in the sun, lose energy. They may fall below escape velocity and be
gravitationally trapped. Trapped WIMPs eventually come to equilibrium and
accumulate near the center of the sun. While the WIMP density builds up, their
annihilation rate into lighter particles increases until equilibrium is achieved where the annihilation rate equals half of the capture rate. The sun
has thus become a reservoir of WIMPs which we expect to annihilate mostly into
heavy quarks and, for the heavier WIMPs, into weak bosons. The leptonic decays of the heavy quark and weak boson annihilation products turn the sun
and earth into nearby sources of high-energy neutrinos with energies in the
GeV to TeV range. Figure~1 displays the neutrino flux from the center of the
earth calculated in the context of supersymmetric dark matter
theories\cite{scopel}. The direct capture rate of the WIMPs in germanium
detectors is shown for comparison. Contours indicate the parameter space
favored by grand unified theories. Most of this parameter space can be covered
by improving the capabilities of existing detectors by 2 orders of magnitude.
Existing neutrino detectors have already excluded fluxes of neutrinos from
the earth's center of order 1~event per $1000 \rm~m^2$ per year. The
best limits have been obtained by the Baksan experiment\cite{suvorova}. They are already excluding relevant parameter space of supersymmetric models. We will
show that, with data already on tape, the AMANDA detector will have an unmatched discovery reach for WIMP masses in excess of 100~GeV.

\begin{figure}[t]
\centering
\hspace{0in}\epsfxsize=4.5in\epsffile{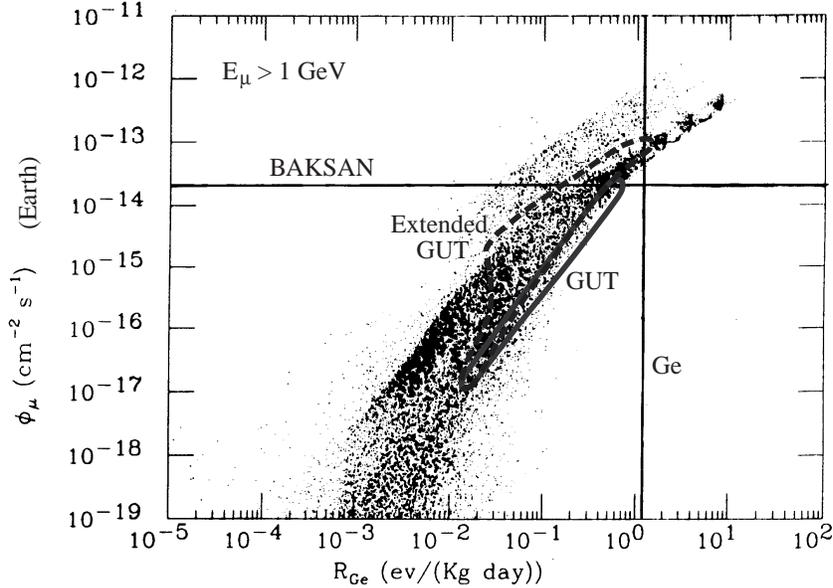}

\caption{Direct and indirect detection rates (for neutrinos from the center
of the earth in the figure shown) of cold dark matter particles predicted by
supersymmetric theory. Grand unified theories favor the parameter space indicated. Part of it is already excluded by present experiments as indicated
by the horizontal and vertical lines.}
\end{figure}

The potential of neutrino telescopes as dark matter detectors has been documented in detail\cite{kamionkowski}. With a sensitivity which increases
with the WIMP mass, they are complementary to direct, cryogenic detectors.
They can detect WIMPS beyond the kinematic limits of the LHC: about 500~GeV
for neutralinos. A striking way to illustrate their potential is to use the
possible detection\cite{belli} in the DAMA NaI detector in the Gran Sasso
tunnel as an example. If their seasonal variation is indeed evidence for
WIMPS, observation of a signal in an exposure of 4500 kg\,days requires a
WIMP-nucleon cross section of $10^{-42}{\sim}10^{-41}$~cm$^2$ for a WIMP mass
of $50{\sim} 150$~GeV. This information is sufficient to calculate their
trapping and annihilation rate in the sun and earth. Both will be a source of,
on average, 100 neutrinos per year of WIMP origin in the existing AMANDA
detector with an effective area of $10^4$~m$^2$. The exact rate varies with
the mass of the WIMP.

\section{Status of the AMANDA Project}

First generation neutrino detectors, launched by the bold decision of the
DUMAND collaboration over 20 years ago to construct such an instrument, are
designed to reach a relatively large telescope area and detection volume for a
neutrino threshold of tens of GeV, not higher. This relatively low threshold
permits calibration of the novel instrument on the known flux of atmospheric
neutrinos.  Its architecture is optimized for reconstructing the Cherenkov
light front radiated by an up-going, neutrino-induced muon. Up-going muons are
to be identified in a background of down-going, cosmic ray muons which are
more than $10^5$ times more frequent for a depth of 1$\sim$2 kilometers. The
method is sketched in Fig.~2.

\begin{figure}[t]
\centering
\hspace{0in}\epsfxsize=2.75in\epsffile{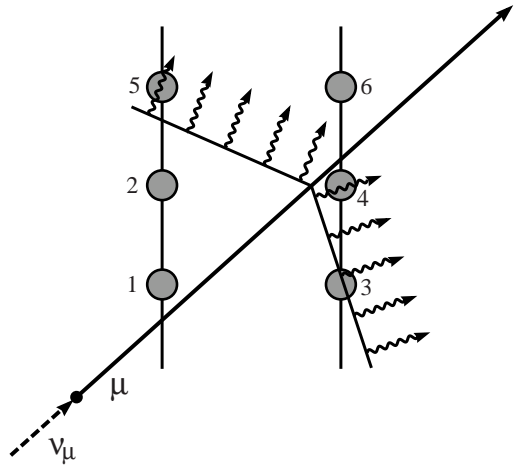}

\caption{The arrival times of the Cherenkov photons in 6 optical modules
determine the direction of the muon track.}
\end{figure}

Construction of the first-generation AMANDA detector\cite{barwick} was completed in the austral summer 96--97. It consists of 300 optical modules
deployed at a depth of 1500--2000~m; see Fig.~3. An optical module (OM) consists of an 8~inch photomultiplier tube and nothing else. OM's have only
failed when the ice refreezes, at a rate of less than 3 percent. Calibration
of this detector is in progress, although data has been taken with 80 OM's
which were deployed one year earlier in order to verify the optical properties
of the ice (AMANDA-80).

\begin{figure}[t]
\centering
\hspace{0in}\epsfxsize=4in%
\epsffile{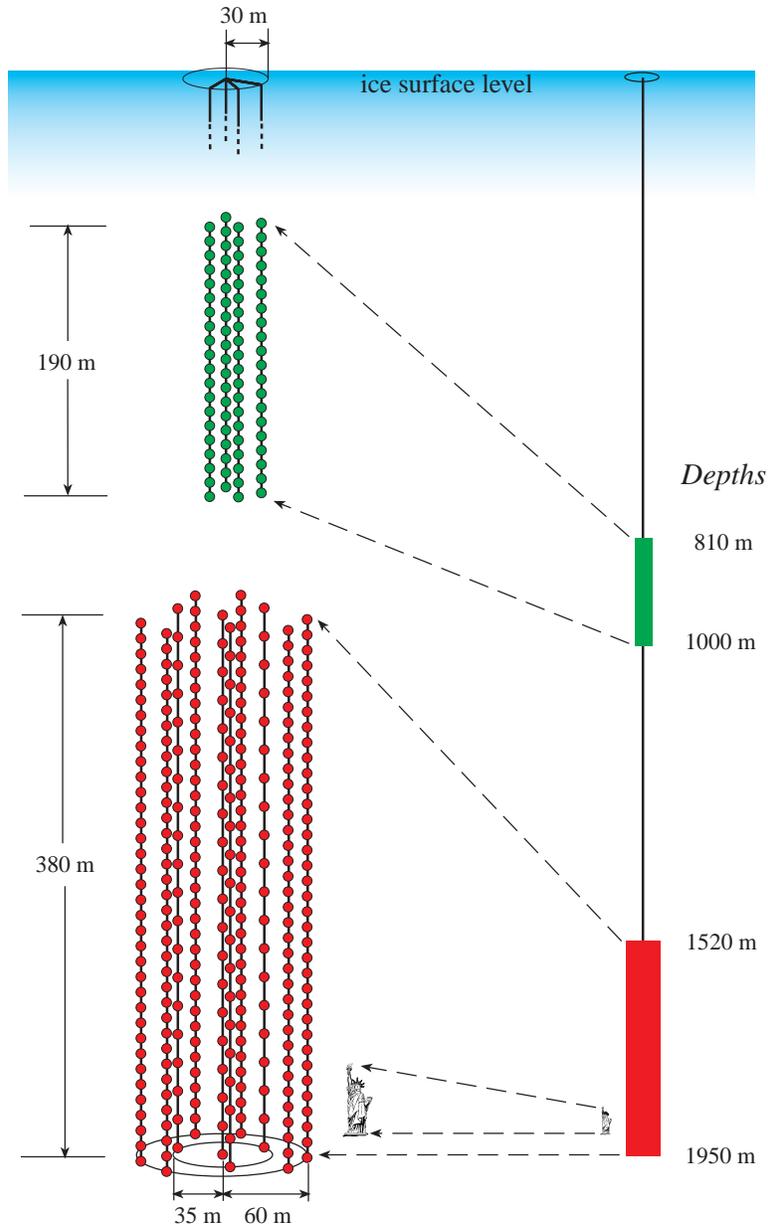}

\caption{The Antarctic Muon And Neutrino Detector Array (AMANDA).}
\end{figure}

As anticipated from transparency measurements performed with the shallow
strings\cite{science} (see Fig.~3), we found that ice is bubble-free at 1400--1500~meters and below. The performance of the AMANDA detector is encapsulated in the event shown in Fig.~4. Coincident events between AMANDA-80
and four shallow strings with 80 OM's have been triggered for one year at a
rate of 0.1~Hz. Every 10 seconds a cosmic ray muon is tracked over 1.2 kilometers. The contrast in detector response between the strings near 1 and
2~km depths is dramatic: while the Cherenkov photons diffuse on remnant bubbles in the shallow ice, a straight track with velocity $c$ is registered
in the deeper ice. The optical quality of the deep ice can be assessed by
viewing the OM signals from a single muon triggering 2 strings separated by
79.5~m; see Fig.~4b. The separation of the photons along the Cherenkov cone is
well over 100~m, yet, despite some evidence of scattering, the speed-of-light
propagation of the track can be readily identified.

\renewcommand{\thefigure}{\arabic{figure}a}
\begin{figure}[t]
\centering
\hspace{0in}\epsfxsize=4.5in\epsffile{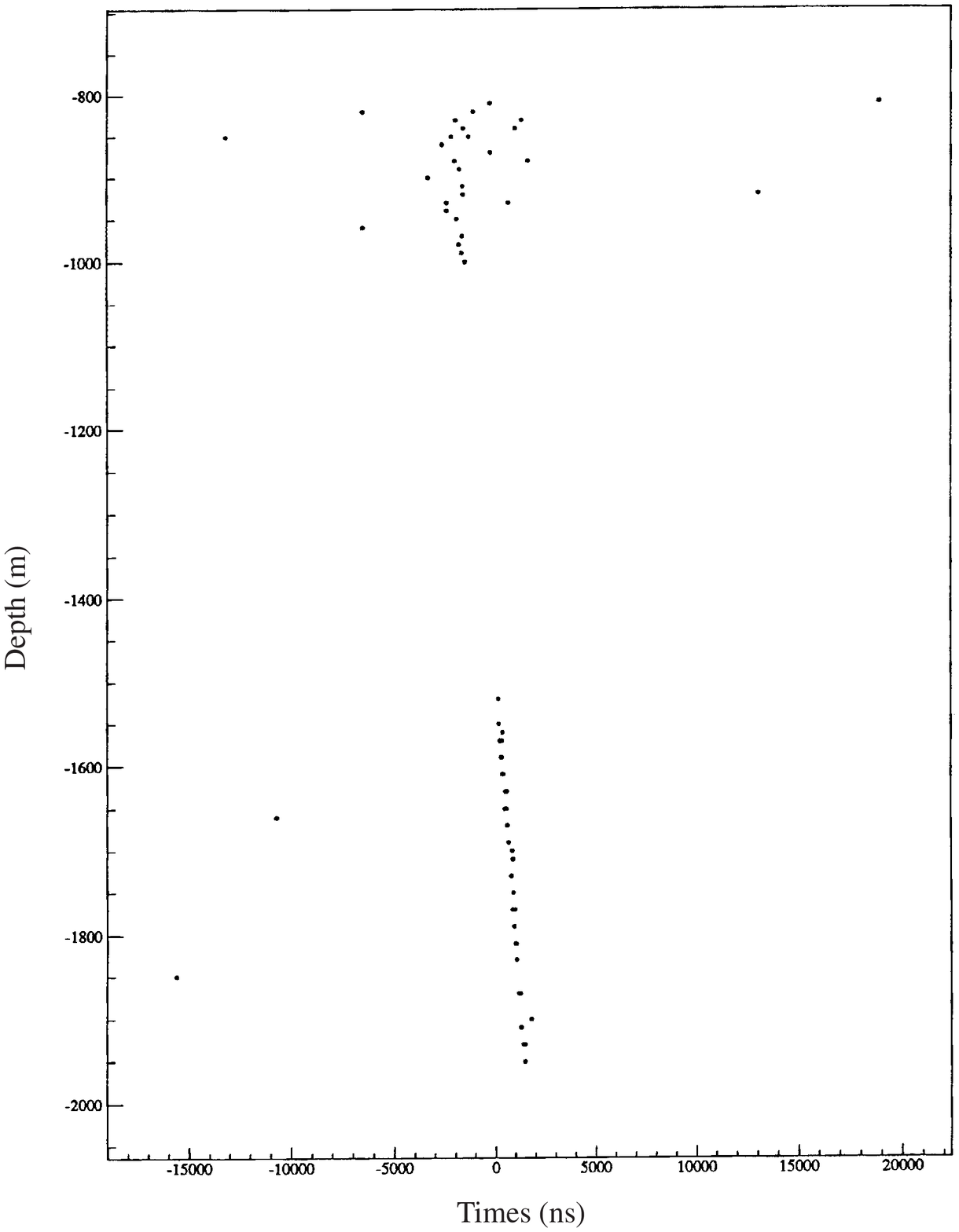}

\caption{Cosmic ray muon track triggered by both shallow and deep AMANDA
OM's. Trigger times of the optical modules are shown as a function of depth.
The diagram shows the diffusion of the track by bubbles above 1~km depth.
Early and late hits, not associated with the track, are photomultiplier noise.}
\end{figure}

\addtocounter{figure}{-1}\renewcommand{\thefigure}{\arabic{figure}b}
\begin{figure}[t]
\centering
\hspace{0in}\epsfxsize=4.5in\epsffile{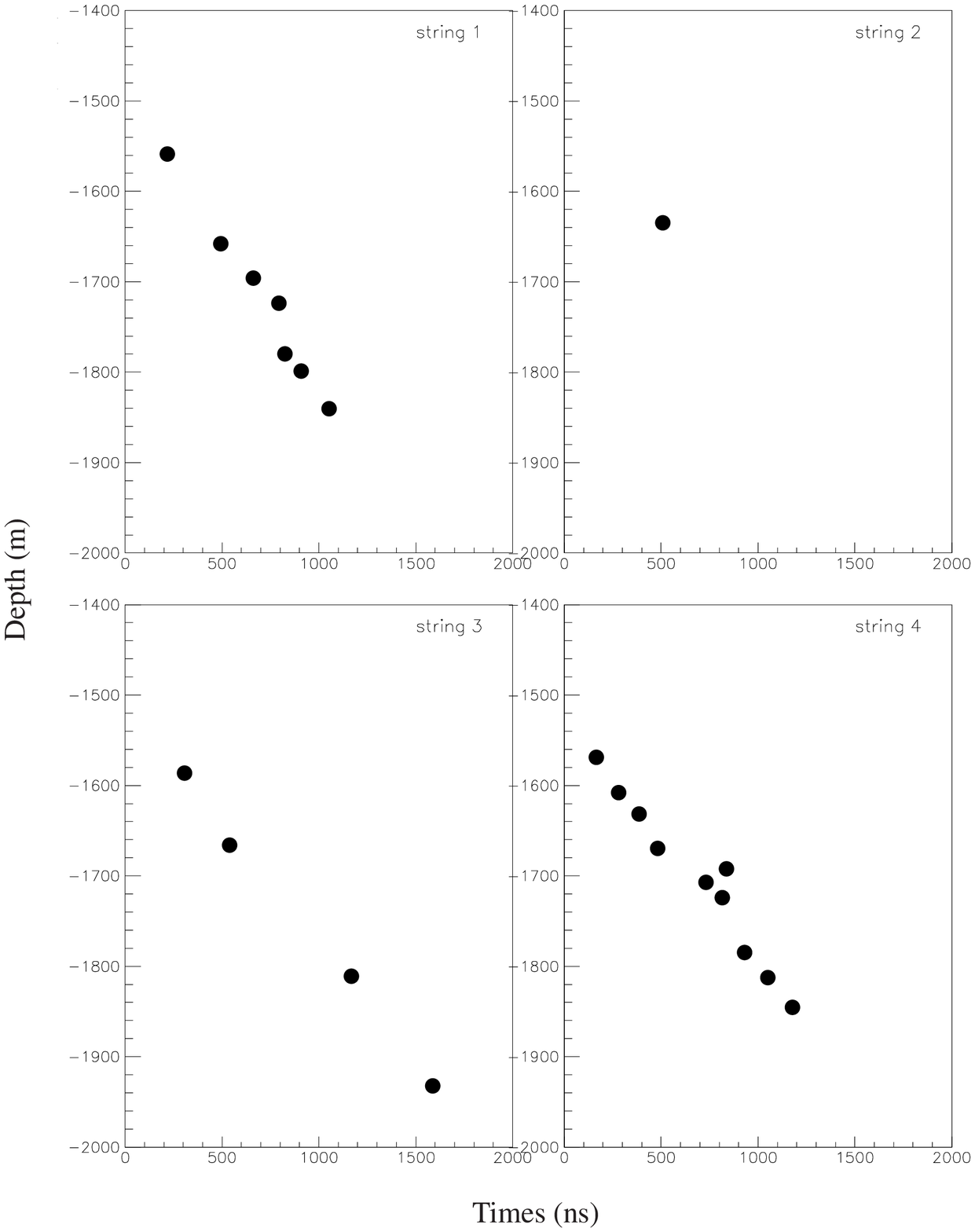}

\caption{Cosmic ray muon track triggered by both shallow and deep AMANDA
OM's. Trigger times are shown separately for each string in the deep detector.
In this event the muon mostly triggers OM's on strings 1 and 4 which are
separated by 79.5~m. }
\end{figure}
\renewcommand{\thefigure}{\arabic{figure}}

The optical properties of the ice are quantified by studying the propagation
in the ice of pulses of laser light of nanosecond duration. The arrival times
of the photons after 20~m and 40~m are shown in Fig.~5 for the shallow and
deep ice\cite{serap}. The distributions have been normalized to equal areas;
in reality, the probability that a photon travels 70~m in the deep ice is
${\sim}10^7$ times larger. There is no diffusion resulting in loss of information on the geometry of the Cherenkov cone in the deep bubble-free ice.

\section{AMANDA: before and after}

The AMANDA detector was antecedently proposed on the premise that inferior
properties of ice as a particle detector with respect to water could be compensated by additional optical modules. The technique was supposed to be a
factor $5 {\sim} 10$ more cost-effective and, therefore, competitive. The
design was based on then current information\cite{dublin}:

\begin{itemize}
\item
the absorption length at 370~nm, the wavelength where photomultipliers are
maximally efficient, had been measured to be 8~m;

\item
the scattering length was unknown;

\item
the AMANDA strategy would have been to use a large number of closely spaced
OM's to overcome the short absorption length. Muon tracks triggering 6 or more
OM's are reconstructed with degree accuracy. Taking data with a simple majority trigger of 6 OM's or more at 100~Hz yields an average effective area
of $10^4$~m$^2$, somewhat smaller for atmospheric neutrinos and significantly
larger for the high energy signals previously discussed.
\end{itemize}

\noindent
The reality is that:
\begin{itemize}
\item
the absorption length is 100~m or more, depending on depth\cite{science};

\item
the scattering length is $25 {\sim} 30$~m (preliminary, this number represents an average value which may include the combined effects of deep ice
and the refrozen ice disturbed by the hot water drilling);

\item
because of the large absorption length, OM spacings are similar, actually
larger, than those of proposed water detectors. Also, a typical event triggers
20 OM's, not 6. Of these more than 5 photons are, on average,
``not scattered\rlap." A precise definition of ``direct" photons will be given
further on. In the end, reconstruction is therefore as before, although additional information can be extracted from scattered photons by minimizing a
likelihood function which matches measured and expected
delays\cite{christopher}.
\end{itemize}

\begin{figure}[t]
\centering
\hspace{0in}\epsfxsize=6in\epsffile{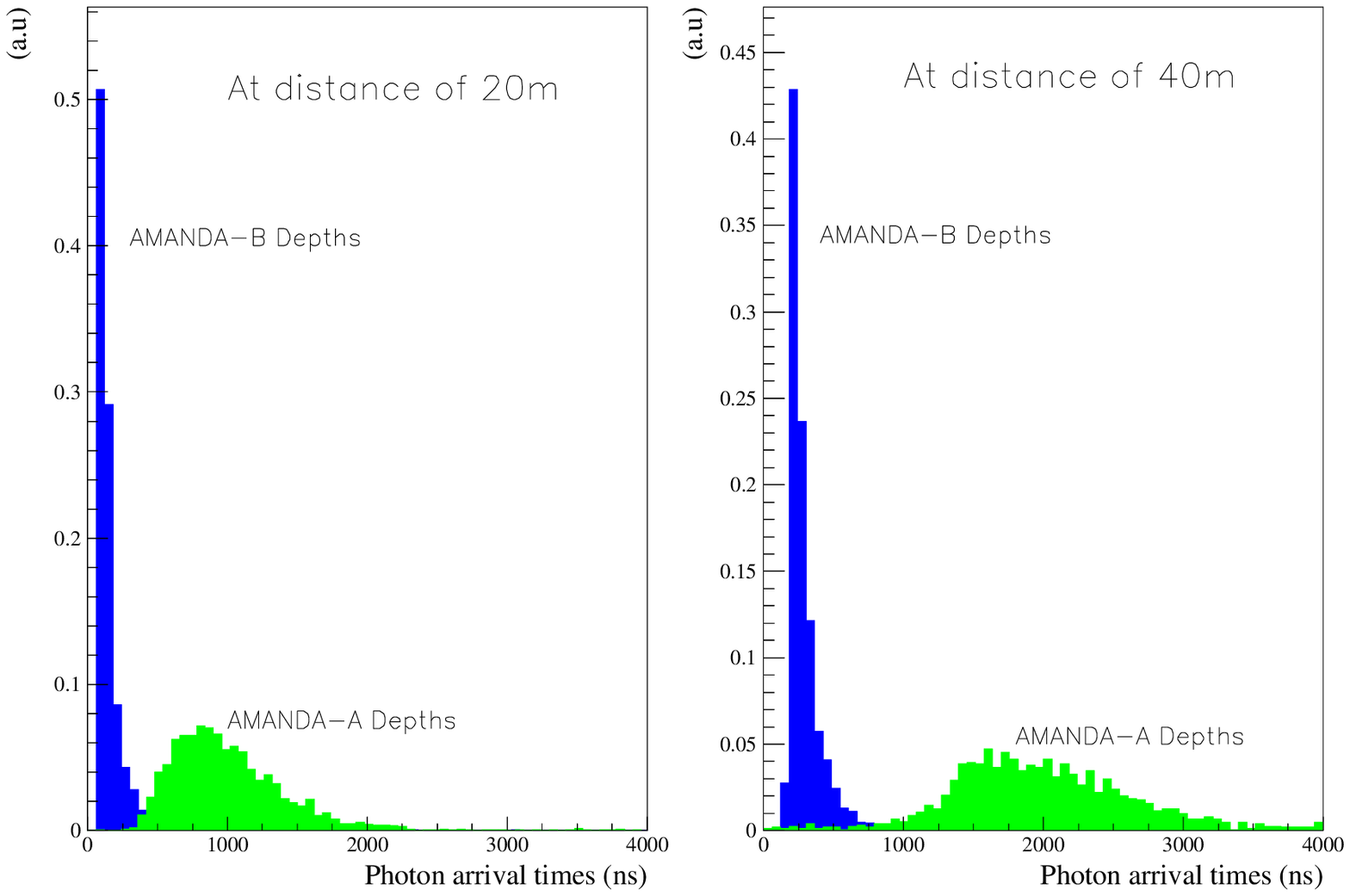}

\caption{Propagation of 510~nm photons indicate bubble-free ice below 1500~m,
in contrast to ice with some remnant bubbles above 1.4~km.}
\end{figure}

The measured arrival directions of background cosmic ray muon tracks, reconstructed with 5 or more unscattered photons, are confronted with their
known angular distribution in Fig.~6. There is an additional cut in Fig.~6
which simply requires that the track, reconstructed from timing information,
actually traces the spatial positions of the OM's in the trigger. The power of
this cut, especially for events distributed over only 4 strings, is very
revealing. It can be shown that, in a kilometer-scale detector, geometrical
track reconstruction using only the positions of triggered OM's is sufficient
to achieve degree accuracy in zenith angle. We conclude from Fig.~6 that the
agreement between data and Monte Carlo simulation is adequate. Less than one
in $10^5$ tracks is misreconstructed as originating below the
detector\cite{serap}. Visual inspection reveals that the remaining misreconstructed tracks are mostly showers, radiated by muons or initiated by
electron neutrinos, misreconstructed as up-going tracks of muon neutrino
origin. At the $10^{-6}$ level of the background, candidate events can be
identified; see Fig.~7. This exercise establishes that AMANDA-80 can be operated as a neutrino detector; misreconstructed showers can be readily
eliminated on the basis of the additional information on the amplitude of OM
signals. Monte Carlo simulation, based on this exercise, confirms that AMANDA-300 is a $10^4$~m$^2$ detector with 2.5 degrees mean angular resolution\cite{christopher}. We have verified the angular resolution of
AMANDA-80 by reconstructing muon tracks registered in coincidence with a
surface air shower array SPASE\cite{miller}. Figure~8 demonstrates that the
zenith angle distribution of the coincident SPASE-AMANDA cosmic ray beam
reconstructed by the surface array is quantitatively reproduced by reconstruction of the muons in AMANDA.

\begin{figure}
\centering
\leavevmode\epsfxsize=6in%
\epsffile{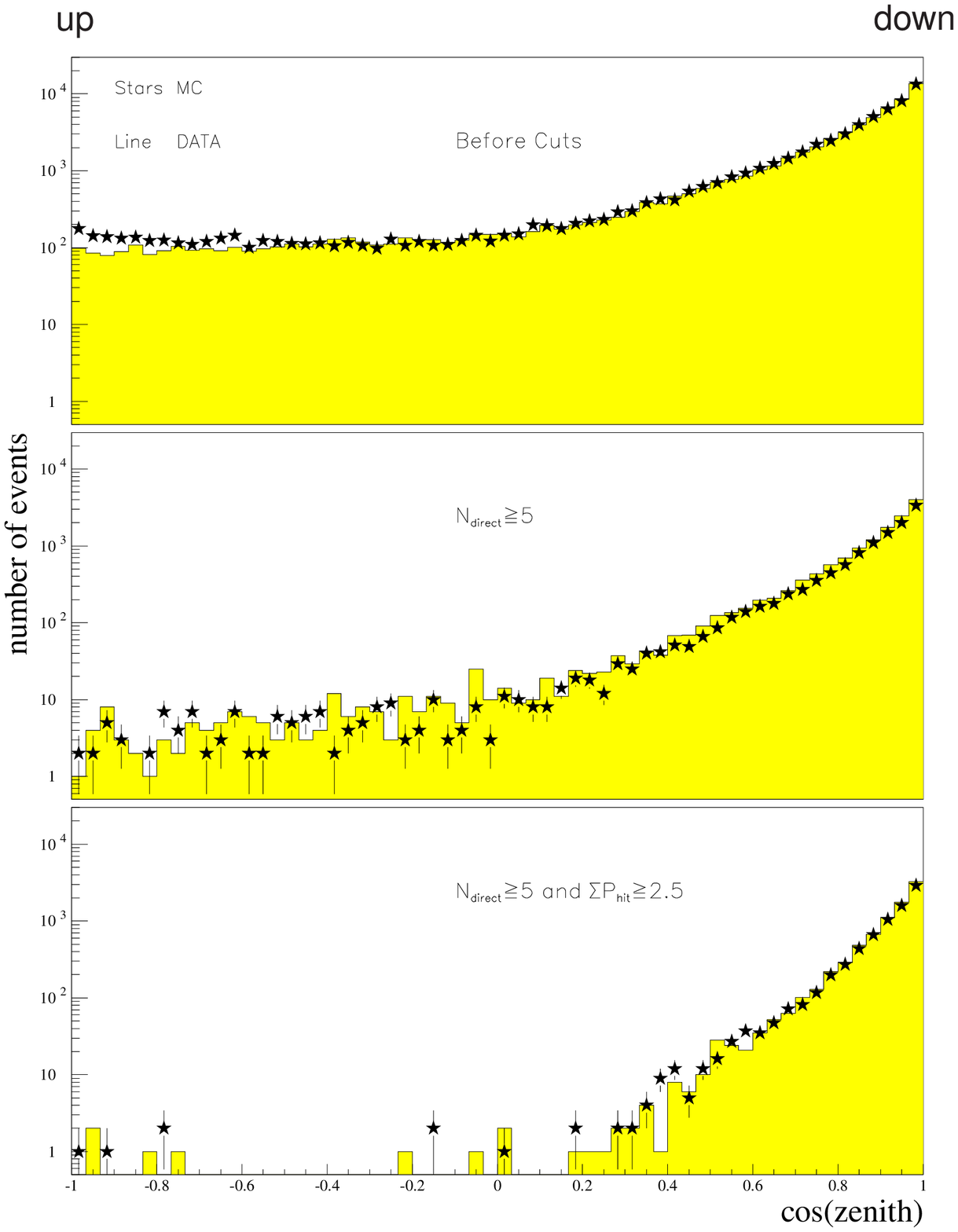}

\caption{Reconstructed zenith angle distribution of muons triggering AMANDA-80: data and Monte Carlo. The relative normalization has not been
adjusted at any level. The plot demonstrates a rejection of cosmic ray muons
at a level of 10$^{-5}$.}
\end{figure}

\begin{figure}[t]
\centering
\hspace{0in}\epsfysize=6in\epsffile{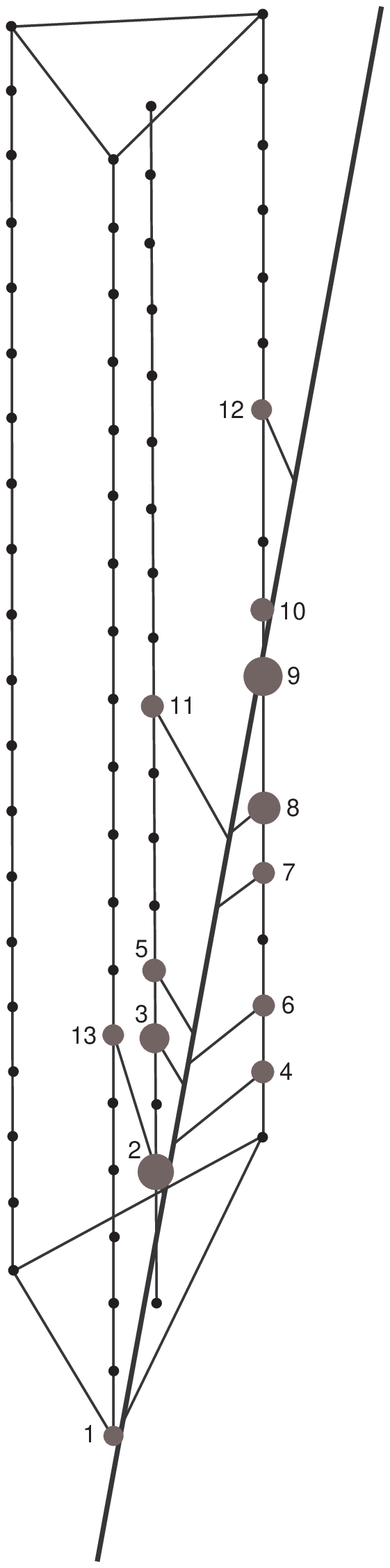}

\caption{A candidate up-going, neutrino-induced muon in the AMANDA-80 data.
The numbers show the time sequence of triggered OMs, the size of the dots the
relative amplitude of the signal.}
\end{figure}

\begin{figure}[t]
\centering\leavevmode
\epsfxsize=2.2in\epsffile{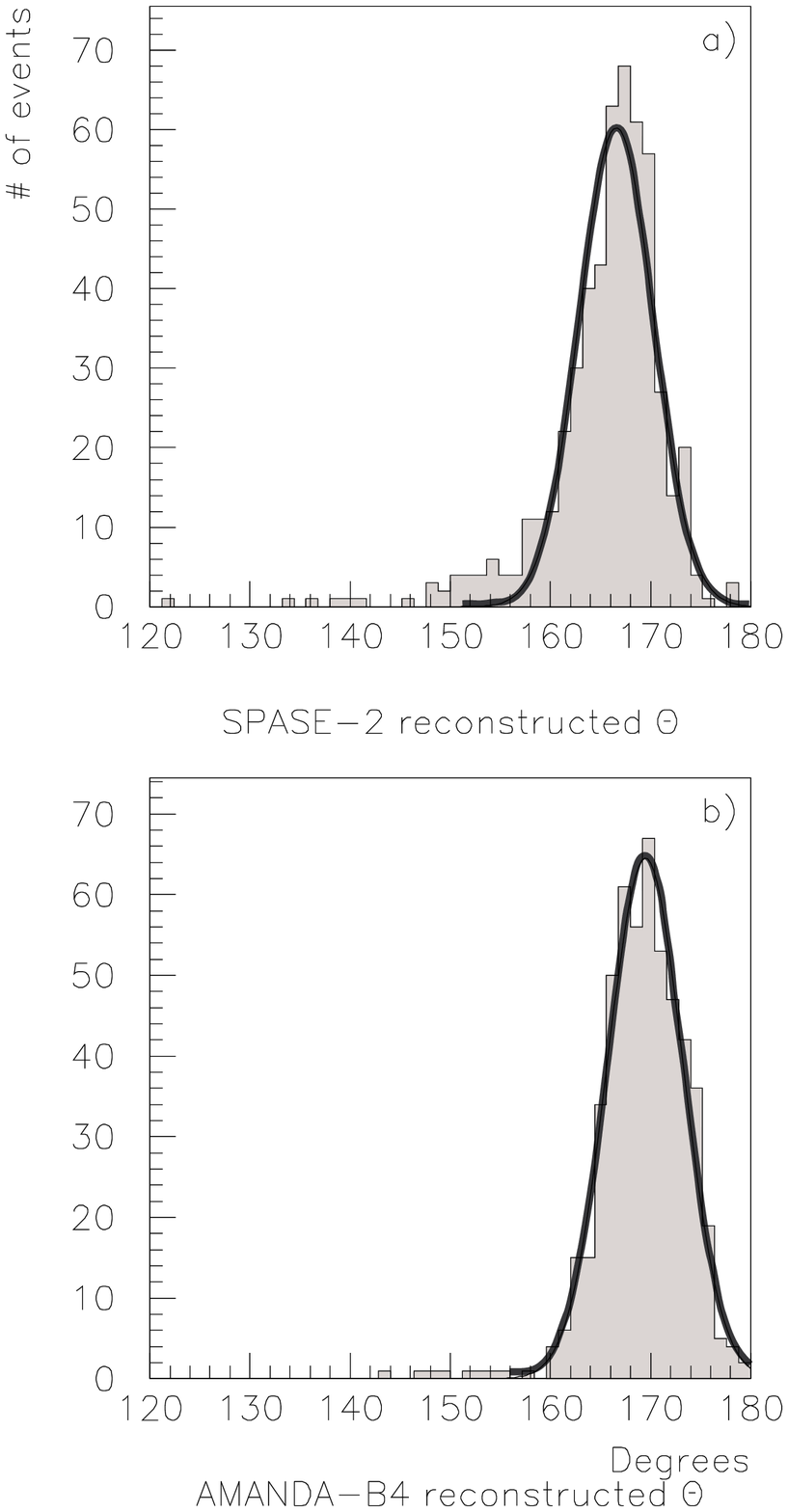}

\caption{Zenith angle distributions of cosmic rays triggering AMANDA and the
surface air shower array SPASE. Reconstruction by AMANDA of underground muons
agrees with the reconstruction of the air shower direction using the scintillator array.}
\end{figure}

\section{Neutrinos from the earth's Center: AMANDA-80}

We have performed a search\cite{bouchta} for upcoming neutrinos from the
center of the earth. One should keep in mind that the preliminary results are
obtained with only 80 OMs, incomplete calibration and only 6 months of data.
We nevertheless obtain limits near the competitive level of less than 1 event
per $250\rm\,m^2$ per year for WIMP masses in excess of 100~GeV. With
calibration of AMANDA-300 completed and 1 year of data on tape, we anticipate
a sensitivity beyond the limit shown in Fig.~1.
Increased sensitivity results from: lower threshold, better calibration (factor of 3), improved angular resolution (factor of $\sim$2), longer exposure and, finally, an effective area larger by over one order of magnitude. Recall that, because the search is limited by atmospheric neutrino
background, sensitivity only grows as the square root of the effective area.

We reconstructed 6 months of filtered AMANDA-80 events subject to the conditions that 8 OMs report a signal in a time window of 2 microseconds.
While the detector accumulated data at a rate of about 20~Hz, filtered events
passed cuts\cite{jacobsen} which indicate that time flows upwards through the
detector. In collider experiments this would be referred to as a level 3
trigger. The narrow, long AMANDA-80 detector (which constitutes the 4 inner
strings of AMANDA-300) thus achieves optimal efficiency for muons pointing
back towards the center of the earth which travel vertically upwards through
the detector. Because of edge effects the efficiency, which is, of course, a
very strong function of detector size, is only a few percent after final cuts,
even in the vertical direction. Nevertheless, we will identify background
atmospheric neutrinos and establish meaningful limits on WIMP fluxes from the
center of the earth.

\begin{figure}[t]   
\centering\leavevmode
\epsfxsize=4.3in\epsffile{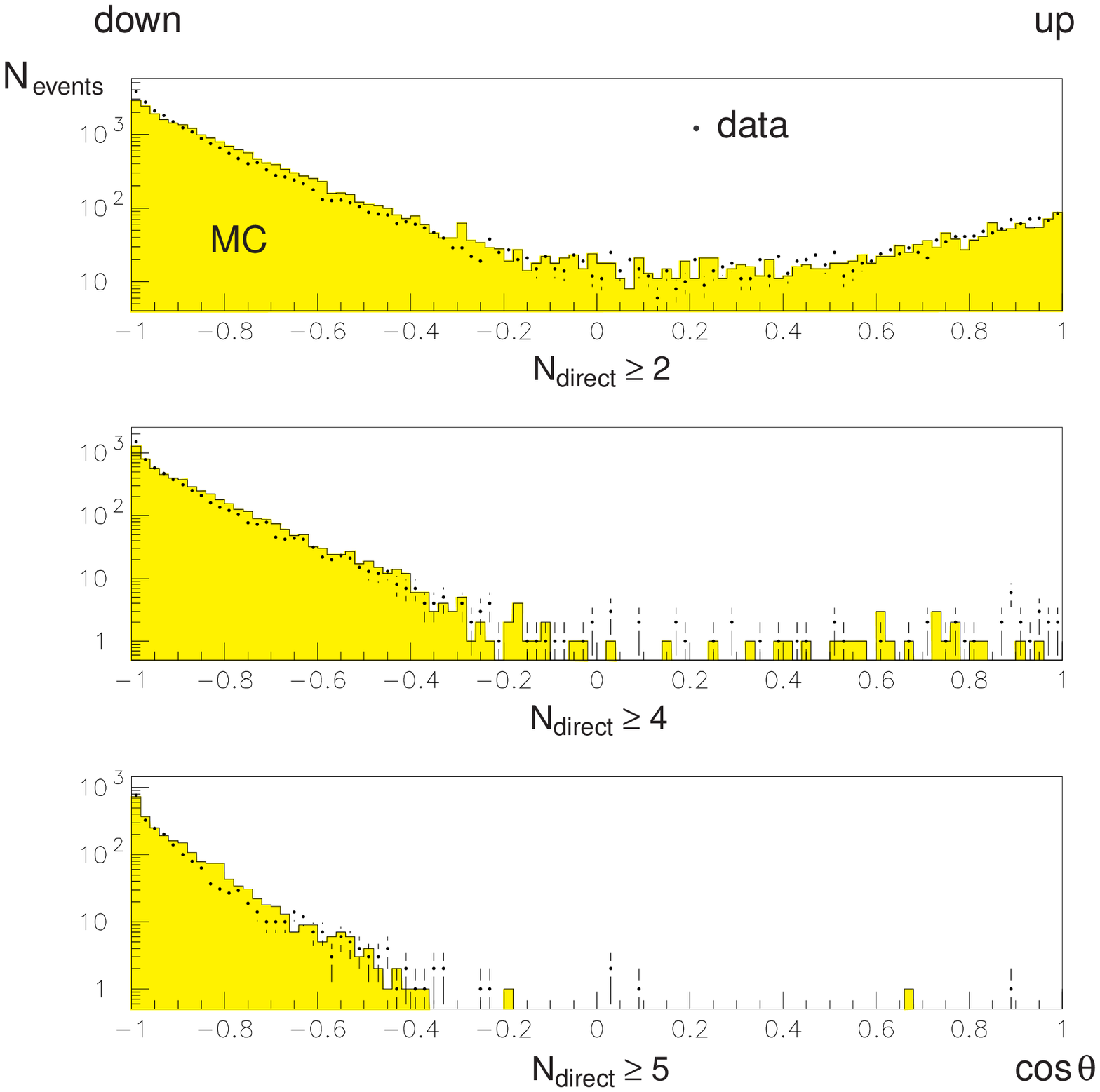}

\caption{$\cos(\theta_{\rm rec})$ is shown with cuts on the number of residuals in the interval $[-15; 25]$~ns. The histogram represents Monte Carlo
simulations with a trigger of 8 or more hits in 2~msec and the dots represent the real data. (Notice that up and down directions are reversed from Fig.~6)}
\end{figure}

That this data set, including prefiltering, is relatively well simulated by
the Monte Carlo is shown in Fig.~9. The results reinforce the conclusions,
first drawn from Figs.~6 and 8, that we understand the performance of the detector. Cuts are on the number of ``direct'' photons, i.e.\ photons which arrive within time residuals of $[-15; 25]$~ns relative to the predicted time. The latter is the time it takes for Cherenkov photons to reach the OM from the reconstructed muon track. The choice of residual reflects the present resolution of our time measurements and allows for delays of slightly scattered photons.
The reconstruction capability of AMANDA-80 is illustrated in Fig.~10. Comparison of the reconstructed zenith angle distribution of atmospheric muons
and the Monte Carlo is shown in Fig.~10a for 3 cuts in $N_{direct}$. For
$N_{direct} \geq 5$, the resolution is 2.2~degrees as shown in Fig.~10b.

\renewcommand{\thefigure}{\arabic{figure}a}
\begin{figure}
\centering\leavevmode
\epsfxsize=3.56in\epsffile{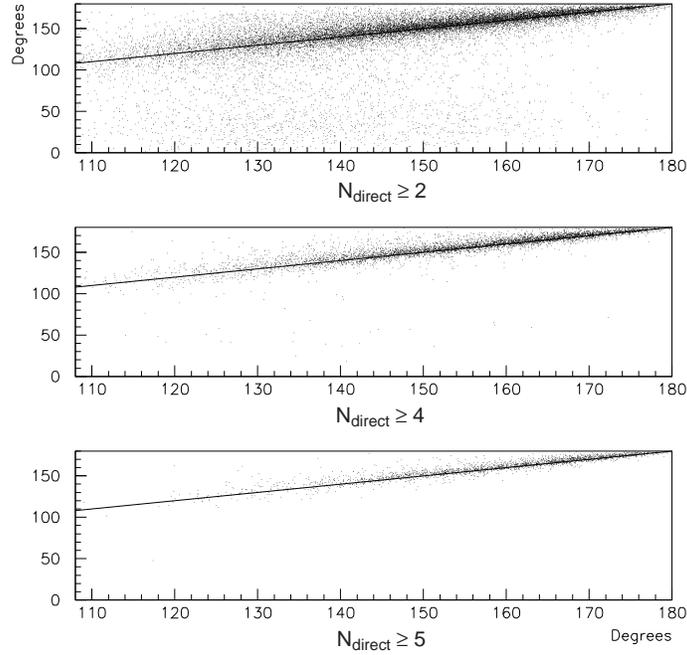}

\caption{Scatter-plot showing the AMANDA-reconstructed $\theta$-angle of
atmospheric muons versus the MC, at several cut levels.}
\end{figure}

\addtocounter{figure}{-1}\renewcommand{\thefigure}{\arabic{figure}b}
\begin{figure}
\centering\leavevmode
\epsfxsize=3.56in\epsffile{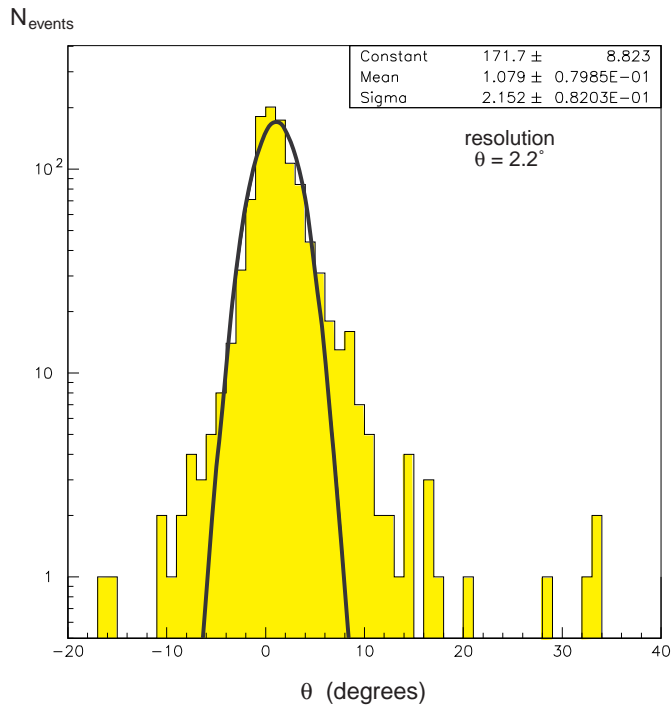}

\caption{$\theta_{\rm rec} - \theta_{\rm MC}$ for reconstructed atmospheric Monte Carlo simulated muons with at least five residuals in the
interval $[-15; 15]$~ns.}
\end{figure}
\renewcommand{\thefigure}{\arabic{figure}}

The final cut selecting WIMP candidates requires 6 or more residuals in the
interval $[-15, +15]$~ns and $\alpha \geq 0.1$~m/ns. Here $\alpha$ is the
slope parameter obtained from a plane wave fit $z_i= \alpha t _i+ \beta$, where $z_i$ are the vertical coordinates of hit OMs and $t_i$ the times at which they were hit. The two
events surviving these cuts are shown in Fig.~11. Their properties are summarized in Table~1. The expected number of atmospheric neutrino events
passing the same cuts is $4.8 \pm0.8 \pm1.1$. With only preliminary calibration, the systematic error in the time-calibration of the PMTs is $\sim$15 ns. This reduces the number of expected events to $2.9 \pm0.6 \pm0.6$. The fact that the parameters of both events are not close to the cuts imposed, reinforces their significance as genuine neutrino candidates. Their large tracklengths suggest neutrino
energies in the vicinity of 100~GeV which implies that the parent neutrino
directions should align with the measured muon track to better than 2~degrees. Conservatively, we conclude that we observe 2 events on a background
of 4.8 atmospheric neutrinos. With standard statistical techniques this result can be converted into an upper limit on an excess flux of WIMP origin; see Fig.~12.

\begin{figure}[t] 
\centering\leavevmode
\epsfysize=6in\epsffile{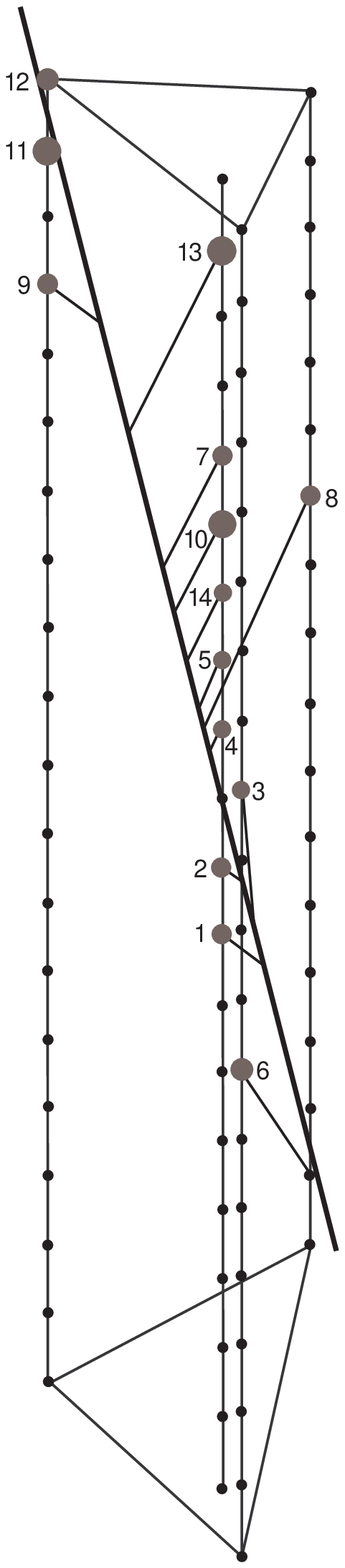}\hspace{1in}
\epsfysize=6in\epsffile{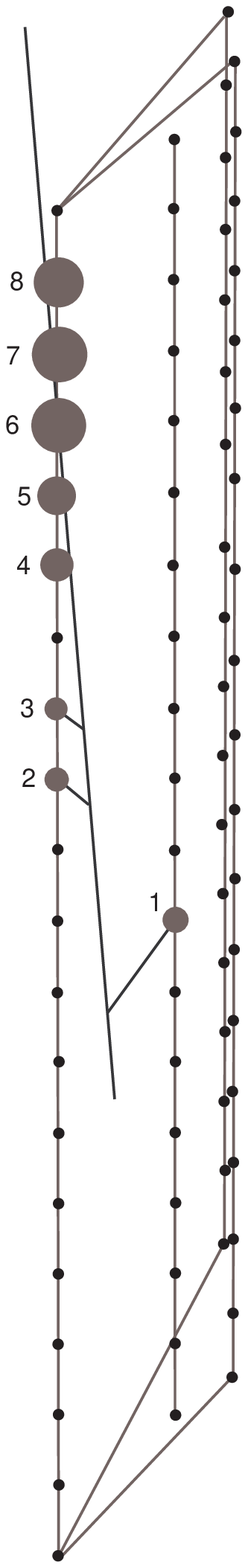}

\caption{Atmospheric neutrino events passing the cuts imposed on the data to
search for neutrinos from WIMP annihilation in the center of the earth.}
\end{figure}

\begin{table}[t]
\begin{center}
\begin{tabular} {|l||c|c|}
\hline
Event ID\# &4706879 & 8427905 \\
\hline
$\alpha$ [m/ns]& 0.19 & 0.37\\
Length [m] &295 & 182\\
Closest approach [m]&2.53  &1.23  \\
$\theta_{rec}$[$^\circ$] &14.1 & 4.6 \\
$\phi_{rec}$[$^\circ$]  &92.0  &  348.7\\
Likelihood/OM &  5.9 &  4.2 \\
OM multiplicity & 14 & 8  \\
String multiplicity & 4  &2  \\
\hline
\end{tabular} \\
\caption{Characteristics of the two events
reconstructed as up-going muons.}
\label{tab:two_events}
\end{center}
\end{table}


In order to interpret this result, we have simulated AMANDA-80 sensitivity to
the 2 dominant WIMP annihilation channels\cite{joakim}:
into $b\bar b$ and $W^+ W^-$. The
upper limits on the WIMP flux are shown in Fig.~12 as a function of the WIMP
mass. Limits below 100~GeV WIMP mass are poor because the neutrino-induced
muons (with typical energy $\simeq m_{\chi}/6$) fall below the AMANDA-80
threshold. For the heavier masses, limits approach the limits set by other
experiments in the vicinity of  $10^{-14}\rm\, cm^{-2}\, s^{-1}$. We have
previously discussed how data, already on tape from AMANDA-300, will make new
incursions into the parameter space of supersymmetric models shown in Fig.~1.

\begin{figure}[t] 
\centering\leavevmode
\epsfxsize=4.5in\epsffile{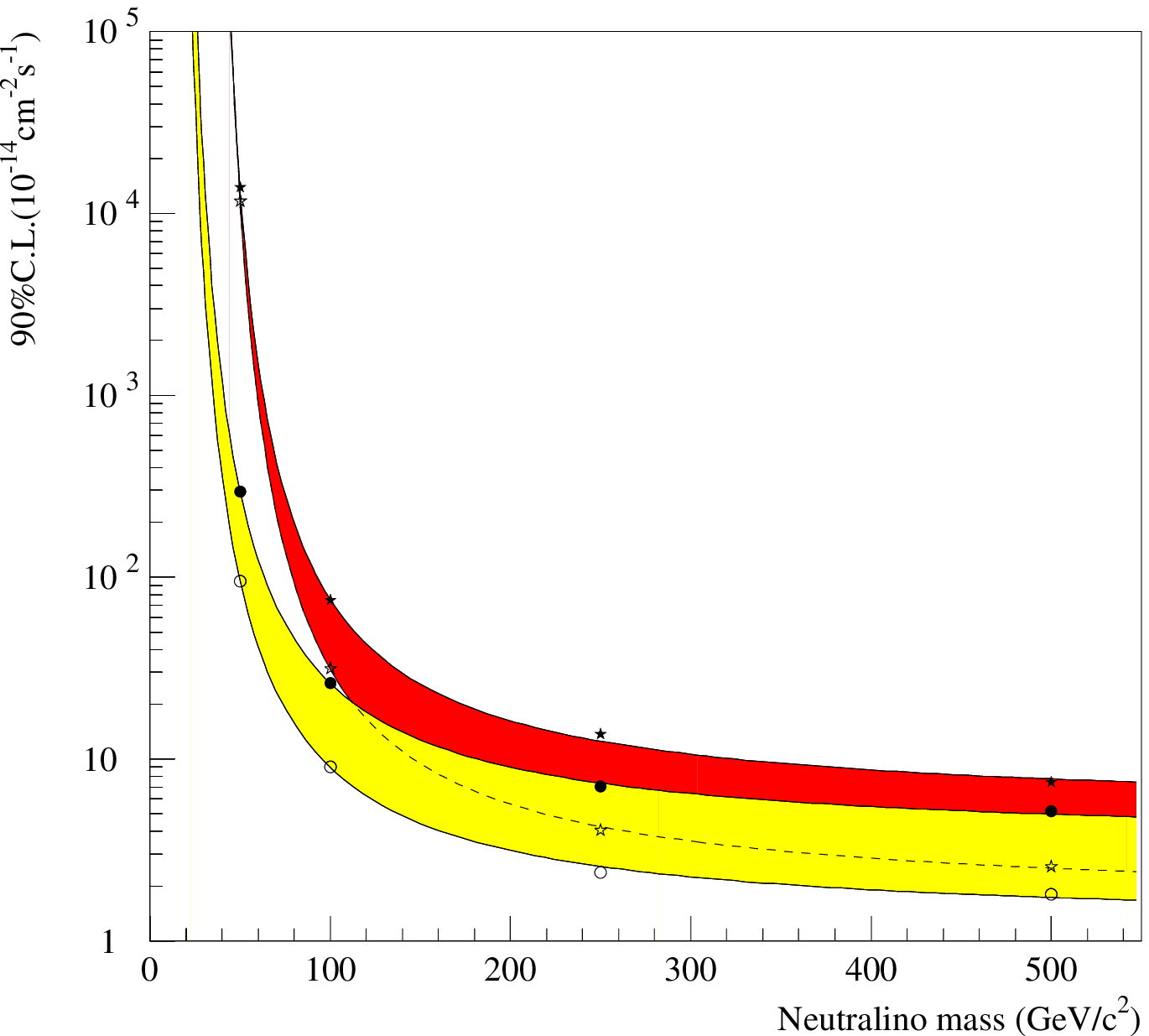}

\caption{Upper limit at the 90\% confidence level on the muon flux from the
center of the earth as a function of neutralino mass.
The light shaded band represents the $W^+ W^-$ annihilation channel
and the dark one represents the $b\bar b$ annihilation channel. The
width of the bands reflects the inadequate preliminary calculation.}
\end{figure}

\section*{Acknowledgements}

The AMANDA collaboration is indebted to the Polar Ice Coring Office and to Bruce Koci for the successful drilling operations, and to the National Science Foundation (USA), the Swedish National Research Council, the K.~A.~Wallenberg Foundation and the Swedish Polar Research Secretariat.
F.H. was supported in part by the U.S.~Department of Energy under
Grant No.~DE-FG02-95ER40896 and in part by the University of Wisconsin Research Committee with funds granted by the Wisconsin Alumni Research Foundation.


\end{document}